# Phase matching of high harmonic generation in the soft and hard X-ray regions of the spectrum

Tenio Popmintchev[1], Ming-Chang Chen[1], Alon Bahabad[1], Michael Gerrity[1], Pavel Sidorenko[2], Oren Cohen[2], Ivan P. Christov[3], Margaret M. Murnane[1*], and Henry C. Kapteyn[1]

[1] JILA, University of Colorado at Boulder, Boulder, CO 80309-0440

[2] Physics Department and Solid State Institute, Technion - Israel Institute of Technology, Haifa 32000, Israel

[3] Department of Physics, Sofia University, Sofia 1164, Bulgaria

*440 UCB, Boulder, CO 80309-0440, Phone: 303-492-7873, FAX: 303-492-5235, murnane@jila.colorado.edu

CLASSIFICATION: PHYSICAL SCIENCES, Applied Physical Sciences

MANUSCRIPT INFORMATION: Number of text pages—31, Number of figures—8

**ABSTRACT**

We show how bright, fully coherent, hard x-ray beams can be generated through nonlinear upconversion of femtosecond laser light. By using longer-wavelength mid-infrared driving lasers of moderate peak intensity, full phase matching of the high harmonic generation process can extend, in theory, into the hard x-ray region of the spectrum. We identify the dominant phase matching mechanism for long wavelength driving lasers, and verify our predictions experimentally by demonstrating phase-matched up-conversion into the soft x-ray region of the spectrum around 330 eV using an extended, high-pressure, gas medium that is weakly ionized by the laser. Scaling of the overall conversion efficiency is surprisingly favorable as the wavelength of the driving laser is increased, making useful, fully coherent, *multi-keV* x-ray sources feasible. Finally, we show that the rapidly decreasing microscopic single-atom yield at longer driving wavelengths is compensated macroscopically by an increasing optimal pressure for phase matching and a rapidly decreasing reabsorption of the generated light at higher photon energies.

**INTRODUCTION**

Advances in x-ray science and technology have resulted in breakthrough discoveries ranging from unraveling the structure of DNA and proteins, to visualizing atoms, molecules and materials at the nanoscale level. These continuing successes have spurred the development of next-generation, ultrafast, coherent, x-ray free-electron laser sources that promise to create super-excited states of matter, or to capture the structure of a single-biomolecule.

Another very exciting advance in x-ray science has been the ability to generate *coherent* x-rays using a tabletop-scale apparatus, by making use of the relatively recent discovery of the extreme nonlinear optics of high-order harmonic generation. Using this process, the output of a tabletop femtosecond laser can be upconverted into the extreme-ultraviolet and soft x-ray regions of the spectrum. The unique characteristics of high harmonic soft x-rays have opened up many new scientific opportunities. The femtosecond-to-attosecond pulse duration has made it possible to capture the coupled motions of electrons, atoms, and molecules in real time (1-7). Moreover, the low divergence and full spatial coherence of the high harmonic light source under the proper generation conditions (8) has also enabled static and dynamic diffraction and imaging with resolutions of tens of nanometers (9-12).

The microscopic physics of the high harmonic generation (HHG) process can be understood in terms of an intuitive three-step semiclassical model (13, 14) (also known as an "atomic antenna"(15)). In this simple model, an electron is first ripped from an atom by the strong electric field of the focused laser light. Once liberated, the electron is accelerated to high energies by the oscillating laser field and can violently recollide with its parent ion. Finally, if the electron recombines with the ion, any excess kinetic energy, acquired in the external field, is emitted in the form of a high-energy photon. This simple physics can be used to determine a maximum (cutoff) photon energy, that can extend from the VUV to the keV regions of the spectrum (16):

$$h\nu_{cutoff} = I_p + 3.17 U_p, \quad \text{where} \quad U_p = 9.33 \times 10^{-14} I\left(W/cm^2\right)\left[\lambda(\mu m)\right]^2 (eV) \quad (1)$$

Here $U_p$ is the ponderomotive energy, or the average kinetic energy of a free electron oscillating in the field of a laser, while $I_p$ is the ionization potential of the atom. It is worth pointing out that this HHG process represents an extreme in nonlinear optics. The energy scale (hundreds of electron-volts to keV in a single atom), time-scale (the x-rays are emitted as a sequence of attosecond ($10^{-16}$-$10^{-18}$ sec) bursts), and spatial scale (Å) of this highly nonlinear frequency conversion process all represent physical extremes.

To date, however, most applications of HHG sources have employed relatively low energy photons, $h\nu \sim <100$ eV. This is because, although each atom emits harmonics over a broad range of photon energies, in order to generate a bright output beam the emission of a large number of atoms over an extended region of the nonlinear gas medium must add in phase (see Fig. 1) (17-20). This *phase matching* condition is met when the driving laser and the generated x-ray waves travel at the same phase velocity, so that the HHG emission from all atoms adds constructively in the forward direction. When photons of energy below ≈ 130 eV are generated, HHG can be perfectly phase matched by balancing the neutral atom dispersion with the dispersion of the free-electron plasma that is created as the medium is ionized, including any geometric contributions (18-20). Under these conditions, HHG emission from cm-long regions can be coherently combined, resulting in an EUV light source with microwatts of coherent power, driven by a compact tabletop laser with average power ~ 1W. This corresponds to an EUV photon flux of $10^{12}$ photons/sec (conversion efficiency ≈ $10^{-5}$) - sufficient for a host of applications in science and technology (1-7, 9-12).

However, from Eq. 1 it is clear that the highest harmonic photon energies are generated at high laser intensities, which strongly ionize the gas medium. The strong dispersion of the resulting free-electron plasma increases the phase velocity of the driving laser so that it is no longer matched to that of the HHG light. Thus, at photon energies >130eV for high harmonic upconversion of 0.8 μm laser light, the HHG process cannot be phase matched using any traditional approach. Without phase matching, coherent signal buildup is limited to the coherence length of the process – that is the distance over which the laser and x-ray waves remain in phase. Coherence lengths <100 μm are typical for HHG at gas pressures of tens of torr, dramatically reducing the HHG output to below that required for most applications. Because the reason for this rapid drop-off in flux for high-harmonic sources is not fundamental to the atomic physics of the HHG process, but rather results from a phase slip problem, the x-ray output could be enhanced by several orders of magnitude at photon energies from 100eV to >> keV if a scheme can be found to correct for this phase slip. As a result, practical approaches for implementing phase matched HHG at high photon energies represent a grand challenge – the high-energy frontier – in the area of nonlinear optics.

To solve the phase matching problem for HHG emission at high photon energies, two successful approaches have been explored to date. First, quasi-phase matching (QPM) schemes either eliminate emission, or correct the phase of HHG emission, from regions of the medium that would otherwise contribute destructively to the nonlinear signal. This can be accomplished using either modulated waveguides or counterpropagating light to create periodic patterns in the laser field amplitude and/or phase, leading to the desired HHG amplitude and phase. Quasi phase matching has been experimentally demonstrated over short interaction lengths (<< mm) up to photon energies of ≈ 300 eV, and in theory can operate at keV photon energies and beyond.

However, by their nature, quasi phase matching schemes are quite complex to implement, requiring the use of multiple laser beams or carefully-engineered geometries.

More recently, calculations and experiments (21) have explored how full phase matching scales with the driving laser wavelength. From Eq. 1 it has long been known that the maximum photon energy that can be generated using HHG (the so-called single-atom cutoff) scales as $I_L \lambda_L^2$, where $I_L$ and $\lambda_L$ are the laser intensity and wavelength. Thus, somewhat counter-intuitively, using a *longer*-wavelength infrared (IR) driving laser allows one to generate *shorter* wavelength high-harmonic light from a single atom at a given laser intensity (22). Recent quantum-mechanical analyses, however, have shown that scaling to shorter wavelengths using IR driving lasers comes at a high cost in terms of photon flux, because the effective brightness of a single-atom emitter scales as $\sim \lambda_L^{-5}$ (23-25). This unfavorable scaling arises mainly due to a combination of the spatial spreading of the recolliding electron wavepacket during the three-step process (which gives a factor of $\lambda_L^{-3}$), and the quadratic increase of the cutoff energy (which results in an additional factor of $\lambda_L^{-2}$ for a fixed energy interval).

Fortunately, by using a longer wavelength driving laser, a given harmonic can be generated at much lower laser intensity, reducing ionization in the nonlinear medium. This dramatically increases the photon energy range over which the HHG process can be phase matched (21). Past work from our group demonstrated that by shifting the driving laser wavelength from 0.8 µm to 1.3 µm, full phase matching over cm distances in Ar extends from 50 eV (0.8 µm) to 100 eV (1.3 µm). More recent experimental work demonstrated enhanced harmonic emission from ~mm regions of a supersonically expanding neutral gas at photon energies around 300 eV (in Ne) and 450 eV (in He) (26). Other theoretical work has suggested that HHG at very high photon energies may be obtained in the so-called nonadiabatic self-phase matching configuration, which relies on very short-duration pulses with a cycle-to-cycle variation of the electric field to extend the phase-matching process (27, 28).

Here, we show that by using mid-infrared driving lasers of moderate intensity ($10^{14}$-$10^{15}$ W/cm$^2$), high harmonic generation presents an experimentally feasible and straightforward route for generating bright, fully coherent, beams up to several keV photon energies. Experimentally, using a driving laser wavelength of $\lambda_L$=1.3 µm, we verify our predictions by demonstrating true phase matched harmonic emission over centimeter distances at high gas pressures (>>1 atm), reaching into the water window region of the spectrum around 330 eV. Moreover, phase-matched conversion in this weakly ionized regime is quite advantageous because the driving laser beam experiences minimal nonlinear distortion, resulting in an excellent spatial coherence of the HHG beam (8, 29). Images of the HHG beam in the water window region of the spectrum indicate a well-directed beam, as expected for phase-matched emission. Moreover, through direct comparison between theory and experiment, we quantitatively test our conceptual understanding of how to extend phase matching of the HHG process using mid-IR lasers (18-20): namely, full phase matching is achieved through a balance between the neutral atom and plasma dispersions in a weakly ionized gas (including any geometric terms). This is similar to the case for phase matching of the HHG process using 0.8 µm lasers. Finally and most importantly, we show that the optimum phase matching conditions scale very favorably well into the multi-keV hard x-ray region of the spectrum. This is because re-absorption of the x-rays by the nonlinear medium is rapidly decreasing with increasing photon energy, so that very high, multi-atmosphere, gas pressures can be used. In addition, nonlinear distortion of the laser beam is low due to the required lower laser intensities for longer driving wavelengths. Hence, the ability to implement full phase matched conversion of laser light to x-rays over macroscopic distances, using a high density medium, can mitigate the otherwise rapidly diminishing $\lambda_L^{-5}$ effective single-atom susceptibility. This new regime of phase matched HHG, using a high-pressure, weakly ionized ($\approx 10^1$-$10^{-3}$ %) gas, contrasts with all other approaches for generating harmonics at high photon energies, which produce very low flux levels (non-phase-matched) from multiply-ionized species at low densities (16, 30-32).

**PHASE MATCHING OF HIGH HARMONIC GENERATION**

A schematic of our experimental setup is illustrated in Figure 1. A driving laser beam originating from a wavelength tunable ultrafast laser - optical parametric amplifier system (see Methods), is focused into a gas-filled hollow waveguide to facilitate near plane-wave propagation. For perfect phase matching of HHG, the driving laser phase velocity must equal that of the generated x-rays. In any HHG geometry, the phase mismatch $\Delta k$ is a sum of contributions from the pressure-dependent neutral atom and free electron dispersion, as well as from pressure-independent geometric dispersion, which for the case of a hollow waveguide is given by (18, 19):

$$\Delta k = k_{x-ray} - qk_L \approx q\left( \underbrace{\frac{u_{11}^2 \lambda_L}{4\pi a^2}}_{geometric} - \underbrace{p(1-\eta)\frac{2\pi}{\lambda_L}\left(\Delta\delta(\lambda_L) + \tilde{n}_2 I_L\right)}_{atoms} + \underbrace{p\eta N_{atm} r_e \lambda_L}_{free\ electrons} \right). \quad (2)$$

Here $q$ denotes the harmonic order, $u_{11}$ is the mode factor, $a$ is the inner radius of the hollow waveguide, $p$ is the pressure, $\eta$ is the ionization level, $r_e$ is the classical electron radius, $N_{atm}$ is the number density of atoms per atm, $\Delta\delta(\lambda_L)$ is the difference between the indices of refraction of the gas per atm at the fundamental and x-ray wavelengths, and $\tilde{n}_2$ is the nonlinear index of refraction per atm at $\lambda_L$. Phase matching, i.e. $\Delta k \rightarrow 0$, can be achieved by varying the gas pressure inside the waveguide since the sign of the neutral atom contribution to the dispersion is opposite that of the generated free-electron plasma. This dispersion balance mechanism has recently been directly verified through *in situ* measurements of the coherence length of the HHG process as the phase matching conditions were varied (33).

From Eq. 2, phase matching is possible only if the ionization $\eta$ is less than a critical ionization level given by $\eta_{CR}(\lambda_L) = \left[\lambda_L^2 N_{atm} r_e / (2\pi\Delta\delta(\lambda_L)) + 1\right]^{-1}$ (18-20). Values for $\eta_{cr}$ are on the order of a few percent in the near-IR region, e.g. approximately 4% (1.5%) for Ar, 1% (0.4%) for Ne, and 0.5% (0.2%) for He at 0.8μm (1.3μm) driving laser wavelengths. This critical ionization level monotonically decreases as the driving laser wavelength increases into the mid-IR.

Under the illumination conditions we consider here (laser intensities of $10^{14}$-$10^{15}$ W/cm$^2$ and 8-cycle laser pulses), ionization of an atom by an intense laser pulse is well-described by the Ammosov-Delone-Krainov (ADK) tunneling ionization model, particularly when using longer-wavelength driving laser (34). Using the ADK model, we can estimate the laser intensity for which ionization in the medium approaches $\eta_{CR}$, which from Eq. 1, defines a *phase matching cutoff photon energy* $h\nu_{PM}(\lambda_L) \approx 3.17 U_p \propto I_L(\eta_{cr})\lambda_L^2$ (neglecting $I_p$) (21). This phase matching cutoff $h\nu_{PM}$ corresponds to the maximum photon energy that can be generated from a macroscopic medium with near-optimum conversion efficiency (full phase matching). Figure 2A plots the phase matching cutoff $h\nu_{PM}(\lambda_L)$ for values of $\lambda_L$ up to 10 μm, assuming a hyperbolic-secant laser pulse with 8 optical cycles FWHM (21 fs at $\lambda_L$=0.8μm; 35 fs at $\lambda_L$=1.3 μm). This plot shows that phase matching of HHG can extend to 1 keV for driving laser wavelengths approaching 3 μm, and extends even to the multi-keV x-ray region when longer mid-IR laser wavelengths are used. Use of a shorter 3-cycle pulse (FWHM) can increase these phase matching cutoffs by an additional 15%, due to decreased ionization levels for shorter laser pulses. Finally, phase matching cutoffs may increase by an additional few percent due to non-adiabatic effects, which also lower the ionization level and which are not captured by the quasi-static ADK approximation (35).

**EXPERIMENTAL RESULTS - PHASE MATCHING OVER EXTENDED DISTANCES AT HIGH PRESSURES IN THE WATER WINDOW**

In order to experimentally verify the predicted scaling of the HHG phase matching cutoffs with driving laser wavelength, we generated driving laser beams either from an ultrafast Ti:Sapphire laser amplifier ($\lambda_L$=0.8 μm), or from an optical parametric amplifier, tuned to $\lambda_L$=1.3μm, with energy of up to 5.5 mJ and with a pulse duration under 35 fs. The driving laser was focused into a hollow capillary filled with Ar, Ne or He gas (see Methods). Harmonics generated using 0.8 μm driving beams serve as a reference. At this wavelength, the phase matching cutoff extends to ~50eV, 90eV, and 130eV in Ar, Ne, and He, respectively (18-20, 36). Figure 2B illustrates a typical 2D plot of the observed HHG spectrum as a function of pressure (Ar, $\lambda_L$ =0.8 μm). As shown on a linear scale, the HHG spectrum extends to approximately 50 eV. Figures 2C, D, and E show equivalent pressure-tuned phase matching spectra using a 1.3 μm driving laser. The phase matching cutoffs in Ar (21) and Ne extend to ≈ 100 eV and 200 eV respectively, while for He phase matching extends into the water window region of the soft x-ray spectrum, to ≈ 330 eV. These phase matching cutoff values are all well beyond what can be achieved using a 0.8 μm driving laser. Moreover, in all cases the observed cutoffs (for different wavelengths of $\lambda_L$ =0.8 and 1.3 μm, and various nonlinear media - He, Ne, and Ar) are in good agreement with theoretical predictions of phase matching from Eq. 2. The optimum peak focused laser intensity on-target also agrees well with our calculations.

Additional comparisons between theory and experiment can be made by comparing the experimental and predicted phase matching pressures. The pressure scaling with driving laser wavelength is very favorable. The phase mismatch in Ar driven by 0.8 μm light in a 250 μm waveguide is minimized ($\Delta k \rightarrow 0$) at an optimal pressure of 20-40 torr (Fig.2B) (33). Remarkably, the experimentally observed optimal phase-matching pressure increases from ~30 torr (0.8 μm) to 100-200 torr (1.3 μm) for Ar, from ~100 torr (0.8 μm) to 600-900 torr (1.3 μm) for Ne, and from ~500 torr (0.8 μm) to >2200 torr (1.3 μm) for He (Fig.3 A). Such high pressures are convenient to achieve using a waveguide geometry, where the backing pressures are close to the pressure in the interaction region. On the other hand, in a gas jet the density can be orders of magnitude lower in the interaction region than in the backing pressure region. Plots of harmonic intensity versus pressure in Figure 3B, calculated using a simple semi-analytical model (see discussion in the following section), predict that higher optimal pressure is needed as the laser wavelength increases from 0.8 to 1.3 μm. This calculation was done for photon energies close to the phase matching cutoffs $h\nu_{PM}$ (i.e. harmonics generated in a narrow time window at an ionization level slightly below $\eta_{CR}$). More realistic calculations of the pressure dependence of the HHG signal near the phase matching cutoff, using the strong field approximation (SFA) (37) also confirm that the optimal pressure for phase matching increases with the driving laser wavelength $\lambda_L$ (Fig.3 C). Thus, experiment and theory show very good agreement for the pressures required for full phase matching.

In our numerical simulations, we include the dispersion of the neutral atoms and transient plasma, geometric dispersion, ohmic power dissipation due to ionization, and reabsorption of the harmonics by the medium. Phase matching of the HHG process was determined numerically by superimposing the driving electric field at consecutive positions along the waveguide, calculated in a frame moving at the speed of light in vacuum, *c* (Fig.3 D). In this picture, the dispersion of the neutral atoms and the waveguide results in a phase variation at the leading edge of the pulse, while the intrinsic plasma formation leads to a phase variation at the trailing edge of the laser pulse. Since the highest photon energies are generated only during the most intense central half-cycle of the laser pulse, and propagate at phase velocity of approximately *c*, phase matched emission at the cutoff manifests itself in an absence of phase slip within this peak half-cycle of the laser (Fig.3 D, top inset).

Figure 4 compares the experimentally measured relative phase-matched HHG flux for various gases using a driving laser wavelength of 1.3 μm. The emission from Ar is ~1.5x higher than from Ne when optimized, for propagation lengths close to the absorption limited medium length of about 1 cm (150 torr) and 0.5 cm (700 torr) respectively (38) (note that different thin-film filters are used to block the fundamental light). The emission in He, although not fully optimized because of the very high gas load on the vacuum system, is only 8x lower than that from Ne for the same propagation length. Thus, although the single atom effective high order nonlinear susceptibility for HHG emission from He is much lower than for Ne, the higher relative transparency of He allows larger density-length products to be employed for the conversion medium. Both the single-atom HHG emission strength and the gas transparency are directly related to the photoionization cross section of the atom.

We also note that from Fig. 2, $\lambda_L \approx 1.3$ μm is the shortest driving laser wavelength that can be used to fully phase match HHG emission from He in the water window above the carbon absorption edge at 284 eV. This driving laser wavelength is likely to be optimal for generating bright harmonics in this photon energy region because of the rapidly diminishing $\lambda_L^{-5}$ effective single-atom susceptibility (for Ne the optimal wavelength would be ≈1.6 μm for generating water window HHG photons). Plotted on a logarithmic scale, these HHG spectra (obtained at the optimal pressures) exhibit several characteristics of phase-matched HHG emission – a well defined plateau region followed by a sharp cutoff, good agreement between the measured and calculated phase matching pressures (Fig. 3), and finally, a well defined, low-divergence (<mrad), HHG beam profile, even for very high gas pressures >2200 torr (Fig. 4B). The experimentally measured soft-ray beam generated under phase matching conditions in He (Fig 4B) fits very well to a Gaussian profile, suggesting high spatial coherence (8, 39). At low gas pressure and high laser intensities in a non-phase matched regime, we find that HHG emission using $\lambda_L$=1.3 µm extends to 200 eV, 400 eV and 500 eV in Ar, Ne and He, respectively. However, a much reduced HHG flux is observed in this spectral region above the phase-matching cutoff.

**SCALING OF PHASE MATCHING TO KEV PHOTON ENERGIES**

Since the experimental results presented above agree very well with our theoretical predictions for phase matched HHG driven by IR laser fields, it is useful to predict how phase matching scales to even higher photon energies. We first determine how the critical ionization level $\eta_{CR}$ and the phase matching cutoffs scale with driving laser wavelength $\lambda_L$. Figure 5A shows that the critical ionization level continues to decrease to 0.1-0.001% for driving lasers in the mid-IR wavelength range. Since the index of refraction at the fundamental driving laser wavelength ($\Delta\delta(\lambda_L)$ in Eq. 2) changes only by 0.35% in the range of $\lambda_L$=0.8-10 µm, and since the index of refraction at x-ray wavelengths is ~1 (i.e. $\Delta\delta(\lambda_{x-ray})$~0), wavelength scaling of the critical ionization can be approximated as $\eta_{CR} \propto \lambda_L^{-2}$. ADK ionization rates are wavelength independent and strongly dependent on the laser intensity. Thus, the maximum driving laser intensity that ionizes to the critical ionization level decreases slightly from $\sim 10^{15}$ to $\sim 10^{14}$ W/cm$^2$ as one moves to longer driving laser wavelength (Fig.5 B). As a result, the phase matching cutoff photon energy effectively scales as approximately $h\nu_{PM} \propto \lambda_L^{1.6-1.7}$ (see Fig. 2A). This is slightly less that the $\lambda_L^2$ scaling of the single-atom HHG cutoff, but nevertheless represents a strong favorable scaling with driving laser wavelength.

Moreover, the fact that phase matched conversion to a given phase matching cutoff requires *lower* peak laser intensity for longer driving laser wavelength is very encouraging from an experimental point-of-view. The laser intensities required to phase match the HHG process in the keV region of the spectrum are easily accessible using tabletop laser technology. Furthermore, the use of hollow waveguides is advantageous because of the high phase matching pressures required. Furthermore, waveguide geometries such as multilayer-structure waveguides (40), metallic waveguides(41), and structured "photonic crystal" waveguides (42), all become more feasible to use with lower intensity, longer-wavelength laser beams.

It is also instructive to describe the behavior of each of the dispersion contributions in Eq. 2 as the laser wavelength changes while maintaining full phase matching at $h\nu_{PM}$. In this simple analytical model, the geometric (waveguide) dispersion scales linearly with $\lambda_L$, and therefore, the ionization-dependent dispersion terms exhibit the same scaling. In fact, harmonics near the phase matching cutoffs $h\nu_{PM}$ are generated at ionization levels close to $\eta_{CR} \propto \lambda_L^{-2}$. Therefore, for fully phase-matched HHG ($\Delta k = 0$), Eq. 2 predicts that the optimal pressure will scale as $p_{opt} \propto \lambda_L^2$ (almost independent of the gas species for the region of interest). This quadratic growth of the optimal phase matching pressure with driving laser wavelength results in an effective linear dependence on $\lambda_L$ for both the atomic and free electron dispersions. Finally, this increase of the dispersion from the near-IR to the mid-IR results in stronger phase variation for the leading and trailing edge cycles compared to the central cycle, illustrated in Fig. 3D. In practice, this may lead to a simple macroscopic isolation of a single sub-cycle x-ray burst even using a multicycle driving laser pulse, consistent with recent experiments that demonstrate the generation of a single attosecond pulse by means of temporal gating through dynamic phase matching (43, 44).

It is interesting to note that the optimum phase matching pressure and the critical ionization level scale inversely to each other with driving laser wavelength, so that the free electron density, $n_e = \eta_{CR} p_{opt} N_{atm}$, remains approximately constant at $h\nu_{PM}$. This means that the *same number of atoms per unit volume* are being ionized - the prerequisite step for HHG emission - as the wavelength is increased under optimal phase matching conditions at $h\nu_{PM}$. Quantitatively, this free electron density is, for example, $n_e \sim 10^{17}$ cm$^{-3}$ for a fiber with a 250 μm diameter (and drops further for larger waveguide diameter). This corresponds to a plasma frequency cutoff wavelength $>\sim 100$ μm, well outside the region of interest for this work. Thus, even though the longer wavelength laser beams will be more-sensitive to the plasma-created propagation distortions, the medium still remains underdense.

Another potential issue is that at high gas pressures, nonlinear distortion of the driving laser pulses might occur because of self-focusing or temporal or spectral modulation. Fortunately this is not the case (see Methods). Assuming that the nonlinear refractive index does not change significantly with the driving laser wavelength, the required laser power remains below the critical power for self-focusing. Also, at the relatively low laser intensities needed for mid-IR driving lasers, the phase mismatch induced by the nonlinear refractive index is negligible (<2%) compared with that induced by the linear index, and hence, does not modify the phase-matching conditions.

Next, it is important to attempt to predict how the overall flux of phase matched high harmonics scales as we move to longer wavelength drivers and shorter wavelength x-rays. To capture the primary scaling of the macroscopic phase-matched HHG flux, here we use a simple model for the intensity of a single harmonic $dI_q$ per unit interaction area $dA$ and time $dt$ (19, 38):

$$dI_q \propto \frac{\omega_q^2 \rho^2 |s_q|^2}{\alpha_q^2 + \Delta k^2} \left( 1 + e^{-\frac{L}{L_{abs}}} - 2e^{-\frac{L}{2L_{abs}}} \cos \Delta k L \right), \quad (3)$$

where $\rho$ is the gas density, $s_q$ is the amplitude of the single atom response at harmonic frequency of $\omega_q$, $\alpha_q = \rho \sigma_q / 2$ and $\sigma_q$ are the harmonic field absorption coefficient and absorption cross section, and $L$ and $L_{abs} = (\rho \sigma_q)^{-1}$ are the propagation and absorption lengths. In this semi-analytical model, the single atom response $s_q = (1-\eta)d(\omega_q)$ (where $d(\omega_q)$ is the single-atom dipole moment) is determined numerically using the strong field approximation, taking into account up to 8 electron trajectories. In agreement with the comparison reported in Ref. (23), this model was verified to reproduce the $\lambda_L^{-x}$ scaling of single-atom susceptibility (where $x \approx 5$-$6$ for a fixed photon energy bandwidth and at constant laser intensity), that is predicted by more accurate time-dependent Schrodinger equation analyses. Because of the decreasing laser intensity (Fig.5 B) at increasing photon energies along $h\nu_{PM}$ (Fig.2 A), the scaling for the dipole acceleration as a function of $\lambda_L$ exhibits a stronger power dependence that varies slightly depending on the gas species (see Methods): $x \approx 6.4$ considering emission within a fixed bandwidth of $\lambda / \Delta\lambda = 100$ close to $h\nu_{PM}$. Equivalently, if working in a length gauge instead of an acceleration gauge, the power law scaling for the dipole moment is $x \approx 11.5$-$11.2$ (see Methods). If, instead of a fixed fraction bandwidth, emission strength of a single harmonic order is considered, the scaling of this dipole acceleration and dipole moment are $x \approx 9$-$8.4$ and $x \approx 14$-$12.6$ respectively.

The predicted optimum pressure for phase matching HHG emission at the phase matching cutoff $h\nu_{PM}$ is plotted in Figure 6 A. It is interesting to note how different this high-pressure phase matching regime is from previous work - more than 12 atmospheres of He are required for full phase matched conversion to $h\nu_{PM}$ = 1 keV, assuming a waveguide radius of a=125 μm. At longer laser wavelengths and even higher multi-keV photon energies, the optimum pressure may become too high to confine in a differentially pumped geometry. However, differential pumping may no longer be necessary because higher photon energies can penetrate a thicker window. Furthermore, the phase matching pressure can be reduced by increasing the waveguide radius $a$: $p_{opt} \propto (\lambda_L/a)^2$.

In absorptive media and under phase-matching conditions, the intensity of a single harmonic $dI_q$ grows asymptotically to a value that scales as $dI_{qmax} \propto |\omega_q s_q/\sigma_q|^2$ (see Eq. 2). The signal builds to ~90% of its asymptotical value for a medium length of $L_{med} \approx 6L_{abs}$ (19, 38). In general, at very high photon energies, the absorption cross section drops rapidly - thus the higher optimum phase matching pressure does not reduce the length of the medium required to reach significant flux. In this context, the use of helium as the nonlinear medium for very high photon energies is of particular interest. Due to the absence of inner-shell absorption in the photon energy region of interest, He exhibits a qualitatively different scaling of the absorption-limited propagation length ($L_{med} \approx 6L_{abs} \propto \lambda_L^{3.7}$) compared with Ar and Ne, as shown in Figure 6B. For example, at pressures >12 atm ($a$=125 μm) and a phase matching cutoff of $h\nu_{PM}$=1 keV, the absorption limited length $L_{med}$ for He is >50 cm. This macroscopic density-length product will allow for orders of magnitude greater conversion efficiency from laser light to coherent x-rays than has been demonstrated to date in this region.

Linear and nonlinear evolution of the laser field could limit this length however. In this context, we investigated competing limiting factors such as group velocity mismatch between the laser and the x-ray fields, group velocity dispersion, and ionization-induced laser energy loss. The index of refraction and dispersion of neutral helium are relatively low, leading to a phase mismatch equivalent to only a ~500 μm thick fused silica window under the considered conditions at $h\nu_{PM}$=1 keV. However, the neutral atom, plasma and waveguide contributions to the net dispersion $dn/d\lambda_L$ are all negative at the laser wavelengths of interest (see Methods), and therefore cannot be balanced. This leads to a group velocity of the driving laser that is always less than the group velocity of the x-rays - $c$. Thus, the corresponding group velocity walk-off length $L_{GV}$, over which the laser and x-ray pulses accumulate a temporal walk-off of a laser pulse duration, can become shorter than an absorption length (see Methods). Thus, the quadratic growth of the HHG signal with length will be restricted to characteristic distances on the order of $L_{GV}$ (decreasing from ~10 cm to ~1 cm for driving laser wavelengths from 0.8 to 10 μm). The HHG signal still continues to increase for lengths longer than $L_{GV}$, but more slowly than a quadratic dependence. These walk-off effects can likely be ameliorated however by using tapered waveguides with pressure ramps combined with shaping and/or lengthening the driving laser pulse. For medium lengths > $L_{GV}$, temporal reshaping of the envelope of the x-ray bursts will occur, together with the generation of longer trains of x-ray pulses

The ionization losses of the laser are approximately equal to the energy acquired by the free electrons. An ionized electron gains a potential energy of $I_p$ and a quiver energy that can be estimated both from experimental and theoretical photoelectron spectra (23, 45). We estimate that the laser energy loss in He at $\lambda_L$=3 μm for phase matching in the 1 keV region is 0.03 mJ/cm, corresponding to a 3% loss of the required laser energy per absorption length (e.g.

$L_{abs}$=8.7 cm for $a$=125 μm and laser energy of ~8.8 mJ in an 80 fs pulse duration). Use of a pressure gradient in combination with tapered waveguides may be used to maintain a high peak intensity and optimal phase matching conditions over many cm distances, despite ionization losses (46). Also, to some extent the increase in laser energy loss due to ionization is mitigated because for the laser intensities and wavelengths considered here, ionization lies well within the tunneling-ionization regime. In other words, multiphoton-ionization is suppressed, and the energy distribution of the emitted photoelectrons exhibits a spike-like distribution at energies much less than $2U_p$. Moreover, the ratio of the yield of electrons with kinetic energy $>2U_p$ compared with $<2U_p$ will drop more dramatically than $\lambda_L^{-4.4}$ (45) at higher $h\nu_{PM}$ ($I_L$ is not constant and decreases as shown in Fig. 4B).

Finally, the scaling of the resulting macroscopic HHG intensity under optimal phase matching conditions (Fig. 7) suggests that helium is the best medium for extending bright harmonic emission to keV photon energies, while Ne can be used up to 0.87 keV photon energies. Very remarkably, for a fixed emission spectral bandwidth of $\lambda/\Delta\lambda$=100 (which is realistic for imaging experiments), the macroscopic, absorption-limited HHG flux from He does not change significantly up to a phase matching cutoff of $h\nu_{PM}$~1 keV, and may even increase by as much as 2 orders of magnitude in the multi-keV x-ray region (Fig. 7B). These scaling predictions also indicate that Ar is a practical HHG medium only using near-IR driving fields. The qualitatively different behavior of the macroscopic emission in Ne and Ar compared with He is due to the presence of L-shell absorption edges around 870 eV (Ne) and 250 eV (Ar), leading to a strong increase in the absorption cross section. As discussed above, the group velocity mismatch will reduce the HHG flux below the predicted absorption limited signal. This is most relevant for HHG in He for driving laser wavelengths >2 μm, since the absorption limited length in this gas becomes much longer than the group velocity walk-off length at high photon energies. In contrast, in Ne and Ar the absorption limited length is comparable to $L_{GV}$, and the HHG signal growth is mainly restricted by reabsorption. Ultimately, the HHG emission under ideal phase matching conditions can be further increased (approximately linearly) by using either driving lasers with higher repetition rates (tens to hundreds of kHz) (47), and/or by using higher laser energies to increase the area of the interaction region.

**PONDEROMOTIVE EFFECTS**

As the above analysis shows, phase-matched HHG emission using mid-IR driving lasers scales well into the multi-keV regime, while still requiring non-relativistic laser intensities of $10^{14}$-$10^{15}$ W/cm$^2$ (Fig.5 B). In contrast, when using a 0.8 µm driving laser, intensities $>10^{16}$ W/cm$^2$ are required for non-phase-matched harmonic emission in the keV region of the spectrum. At laser intensities of $I_L$=5x10$^{16}$ W/cm$^2$, the magnetic component $\boldsymbol{B_L}$ of the laser electromagnetic field can no longer be neglected, and the associated Lorentz force $\boldsymbol{F}=e(\boldsymbol{E_L}+1/c\ \boldsymbol{v_e}\mathrm{x}\boldsymbol{B_L})$ induces a drift of the returning electron wavepacket along the direction of laser propagation, away from the parent ion. Therefore, even if the phase mismatch is partially compensated, the decrease in recombination probability represents a fundamental limit for efficient single-atom HHG (48).

Using mid-IR driving wavelengths however, the same cycle-averaged kinetic energy of the recolliding electron, corresponding to the same energy harmonics, is achieved using a lower peak electric field that drops as $E_L \propto \lambda_L^{-1}$. Hence the magnetic field is proportionally smaller. The relativistic field strength parameter $\xi=eE_L\lambda_L/(2\pi m_e c^2)$, measuring the ratio between the amplitude of the speed of the quivering classical electron in $E$ field, and the speed of light (49), is less than 15% (Fig. 8A) for the laser parameters required for phase matching up to $\lambda_L$=10 µm (shown in Fig. 5B). Thus, the electron remains nonrelativistic. However, the magnetic field component of the ponderomotive force cannot be neglected since the excursion time for the recolliding electron is also increasing with longer wavelengths. The drift of the center of mass of the returning electron wavepacket, $d \approx c\xi^2/(2\omega_L) \propto \lambda_L^3 E_L^2$, in the direction of laser propagation, starts to exceed the transverse wavepacket spread, $\Delta_z \approx 2^{3/4}(eE_L\hbar)^{1/2}/(\omega_L m_e^{3/4} I_p^{1/4}) \propto \lambda_L E_L^{1/2}/I_p^{1/4}$ (48, 49), for laser wavelengths longer than ~5.9 μm in He, and ~7μm in Ne (points (1) and (2) in Fig.8 B). Above these laser wavelengths, corresponding to full phase matching cutoffs of about 3 keV (point (1) and (2) in Fig.2 A), the recombination probability, and thus single-atom HHG efficiency, will decrease more rapidly than the drop associated with the 3D spread of the electron wavepacket. The practical upper limit for mid-IR phase matched HHG emission can be more precisely determined through further single-atom calculations that take into account magnetic dipole and electric quadrupole interactions.

**OUTLOOK**

The implications of these scaling predictions are worthy of elaboration. High harmonic generation with mid-IR laser frequencies results in closely separated harmonics due to the lower energy of the fundamental photons. This has the advantage of a nearly continuous spectrum that is semi-continuously tunable, making it suitable for broad applications spectroscopy, approaching synchrotron-based light sources.

In terms of laser technology, longer-wavelength lasers in the mid-infrared region of the spectrum are required. Optical parametric amplification of ultrafast Ti:Sapphire or Yb-based lasers can be used to generate ultrafast pulses throughout the mid-IR wavelength range, while recently-developed laser media that operate in the IR can be employed to directly generate longer wavelengths. Hybrid laser systems may also be very attractive and versatile. The pulse energies, required peak focused intensities, and pulse durations are all very accessible. Thus, a variety of architectures are possible for efficient, robust and practical implementation of this tabletop coherent x-ray source.

The fact that x-ray upconversion occurs at moderate intensity in a weakly ionized gas means that propagation of the beam in the nonlinear medium is likely to be well controlled with minimal nonlinear effects. This fact also makes this regime amenable to the use of more-sophisticated phase-matching techniques that make use of the attosecond time-scale quantum dynamics of the process. For example, in quasi-phase-matching, the phase or amplitude of the high harmonic emission can be periodically modulated to compensate for a finite coherence length in the upconversion process. QPM has been implemented using modulated waveguides(50), or a flexible all-optical alternative – a sequence of counterpropagating light pulses, to periodically turn-off the high harmonic emission (6). New QPM schemes that use a quasi-continuous counterpropagating laser field to obtain grating-assisted phase matching have also been proposed theoretically (51). QPM can be used to selectively phase match a narrow spectral range, resulting in a tunable, nearly-monochromatic ultrafast coherent x-ray source (6). QPM techniques can thus be combined with longer wavelength driving lasers, to extend the region of phase matching beyond the phase matching cutoffs discussed here, while also reducing the required laser intensity to reach a given harmonic order and mitigating magnetic field and nonlinear laser distortion effects.

Another interesting area for future studies is the generation of bright attosecond pulses using mid-IR driving lasers. The relatively rapid variation in phase matching conditions as the driving laser ionizes the medium also raises the possibility that emission will be predominantly in the form of a single sub-cycle burst, as indicated in Fig. 3D. Recent experiments have verified this gated phase-matching effect at photon energies around 45 eV (43, 44).

Finally, the low peak intensity and long driving laser wavelength required to achieve phase matching in the hard x-ray region, along with the long nonlinear medium lengths and high pressures, also make it both amenable and advantageous to use waveguides to confine the laser light and the nonlinear medium. Infrared waveguides coated with metals, metal-dielectric structures, and dielectric layers, or photonic bandgap structures all can work very well to guide infrared laser light, avoid diffraction losses and to phase match the HHG process (52). The lower laser intensity required for IR lasers makes it possible to use these structures, which are limited in the peak intensity damage threshold compared with a simple glass waveguide.

The resulting x-ray source would likely represent a breakthrough in the science of nonlinear optics, where a useful flux of coherent, ultrafast, soft and hard x-rays could be generated from a desktop femtosecond laser source. Applications from magnetic to bio-, nano- and materials imaging and spectroscopy would be enabled. Free-electron laser sources can also benefit tremendously from the availability of these fully coherent x-rays as seeding beams, to improve the spatial and temporal coherence of their output.

**CONCLUSION**

The calculations presented above, verified by experiment in the soft x-ray region of the spectrum, clearly outline a straightforward and feasible route for generating bright, fully coherent, x-ray beams at multi-keV photon energies. By using longer-wavelength driving lasers at moderate, non-relativistic intensities, the mechanism and scaling of macroscopic phase-matching conditions is very favorable. The decreasing effective nonlinearity of the HHG process for longer wavelength driving lasers can be compensated-for because the nonlinear medium is becoming more transparent at higher photon energies, even with the required high optimal gas pressures. For the case of phase-matched HHG emission in He, we predict that the macroscopic absorption limited flux remains constant or increases for higher-energy x-ray generation. Compact, coherent, ultrafast high harmonic sources are already enabling new science in the soft x-ray region. By extending phase matched harmonic emission to the keV photon energy and beyond, many new applications in imaging and spectroscopy will be possible, with spatial and temporal resolution sufficient to capture and control matter at the level of electrons.

**METHODS**

**Driving Light Source** The IR laser source is a 3-stage optical parametric amplifier pumped by a high energy Ti:Sapphire laser with pulse energies of 21 mJ, and a pulse duration of 23 fs, with a high quality spatial profile of $M^2<1.2$, and at a repetition rate of 10 Hz. A white light continuum seed is generated in a sapphire plate and then strongly chirped in time. In three subsequent amplification stages, the signal beam is further amplified in BBO crystals while isolating the idler beam in each of the subsequent steps. The tunability range of the OPA is 1.1-2.8 μm. Conversion efficiencies of 43% (total) were achieved for generating both single and idler beams, with energy of up to 5.5 mJ in the signal beam at a wavelength of 1.3 μm. An output pulse duration sub-35 fs (8 cycles at FWHM at 1.3 μm) was measured using second-harmonic frequency resolved optical gating (SHG FROG).

**Extreme High-Order Harmonic Source and Soft X-ray Detection** The high harmonic beam is generated by focusing an intense driving laser pulse into a gas-filled hollow waveguide. The ratio of the waist of the laser beam to the waveguide radius (a=125 or 200 μm) was ~65%, ensuring guiding of the lowest loss $EH_{11}$ mode. A continuous gas flow, with backing pressures of up to 6000 torr, enters the waveguide through two laser-drilled holes. These holes effectively divide the waveguide into three sections. The end sections with relatively low gas throughput mitigate supersonic expansion of the high-pressure gas, which otherwise would lead to lower pressures in the interaction region. When steady state is established, the static gas pressure along the mid-section has a negligible gradient, and is close to the backing pressure. A flat field soft-X-ray spectrometer and a soft-X-ray CCD are used to detect the generated harmonic beams. Depending on the photon energy range studied, various metal filters (Al, Zr, Ag or Ti) are used to eliminate the fundamental laser light.

**Scaling of Single Atom Yield at the Phase Matching Cutoff**

The power scaling for the single atom yield at the phase matching cutoff $h\nu_{PM}$ is calculated using the strong field approximation (SFA) model, taking into account 8 trajectories. This model reproduced the $\lambda^{-5}$ single-atom effective nonlinear susceptibility scaling at fixed photon energy and for constant laser intensity, as reported in Ref. (23). The predicted photon energies and laser intensities at $h\nu_{PM}$ are shown in Figs. 2A and 5B. Using these parameters, the power law scaling of the HHG yield $P(\lambda_L)$, in a bandwidth $\Delta E$, is calculated using either the Fourier transform of the single atom dipole acceleration $a(\omega)$ or the dipole moment $d(\omega)$:

$$P_a(\lambda_L) = \int_{E-\Delta E/2}^{E+\Delta E/2} |a(\omega)|^2 \, d\omega \propto \lambda_L^x, \text{ and } P_d(\lambda_L) = \int_{E-\Delta E/2}^{E+\Delta E/2} |d(\omega)|^2 \, d\omega \propto \lambda_L^y$$

where $x$ and $y$ are presented in the table below, where $E_f$ is the fundamental laser photon energy, and $E$ – the harmonic photon energy close to the phase matching cutoff. The results are summarized in Table 1:

The interchangeable prefactor for the macroscopically calculated intensity $dI_q$ (Eq. 3) emitted in a specific harmonic at the phase matching cutoff $h\nu_{PM}$, containing all the wavelength scaling terms is given by: $\omega_q^2(\lambda_L) |d_q|^2 \propto \omega_q^{-2}(\lambda_L) |a_q|^2$, where $\omega_q(\lambda_L) \propto \lambda_L^{1.6-1.7}$ is the harmonic frequency at the phase matching cutoff $h\nu_{PM}$. The small variation of the power law scaling of $\omega_q(\lambda_L)$ is a consequence of the dependence of the ionization rate on the ionization potential of the gas (assuming ADK).

**Contribution of the Nonlinear Refractive Index**

The critical power for self-focusing in a waveguide is slightly higher than in the case for a bulk medium, and in general drops with increasing density of the medium: $P_{CR} \approx 1.86\lambda_L^2/(4\pi n_0 p \tilde{n}_2)$, where $n_0$ is the linear refractive index (53). Since the optimal phase matching pressure scales quadratically with laser wavelength, $P_{CR}$ remains constant. On the other hand, the required laser intensity to generate a particular x-ray photon energy decreases for longer wavelength driving lasers. Therefore the laser power always stays below the critical power for catastrophic nonlinear distortion. In the parameter range of interest, the dispersion term related to the nonlinear index of refraction $n_2$ is less than 2% of the linear refractive index term for He, Ne and Ar at $\lambda_L = 0.8$ μm (54). This fraction decreases for mid-IR laser wavelengths. Hence the Kerr nonlinear refractive index does not significantly modify the phase matching conditions (Eq. 2).

**Group Velocity Mismatch**

The group velocity of the driving laser is given by $v_{gL} = c(n_L - \lambda_L dn_L/d\lambda)^{-1}$, where $n_L = 1$ under phase matching conditions. In contrast to phase velocity matching, group velocity matching cannot be achieved since all the terms in $dn_L/d\lambda$ have the same sign:

$$\frac{1}{v_{gL}} - \frac{1}{c} = -\frac{\lambda_L}{c}\frac{dn_L}{d\lambda}\bigg|_{\lambda_L} \approx \frac{\lambda_L}{c}\left(-p(1-\eta)\frac{d\delta(\lambda_L)}{d\lambda}\bigg|_{\lambda_L} + \frac{p\eta N_{atm}r_e\lambda_L}{\pi} + \frac{u_{11}^2\lambda_L}{4\pi^2 a^2}\right)$$

Using waveguides with larger radius ($a$) minimizes $dn_L/d\lambda$ since it scales approximately as $a^{-2}$ at the phase matching cutoff. This increases the group velocity walk-off length ($L_{GV}$) between the driving pulse and an x-ray pulse (corresponding to a temporal walk-off of a laser pulse duration $\tau_{FWHM}$, $L_{GV}=\tau_{FWHM}/(1/v_{gL}-1/c)$). However, increasing the waveguide radius does not lead to a larger ratio of $L_{GV}/L_{abs}$, that is required for the best HHG flux (see Eq. 3), since both lengths scale effectively as $\propto a^2$. Thus, the quadratic growth region for HHG emission is effectively limited to distances comparable with $L_{GV}$.

**ACKNOWLEDGEMENTS**

The authors acknowledge Sterling Backus, Xiaoshi Zhang, and Greg Taft for valuable technical help. This project is funded by the National Science Foundation and the Department of Energy. IPC also acknowledges support from the Bulgarian National Science Foundation.

**FIGURE LEGENDS**

**Figure 1:** Schematic of extreme nonlinear upconversion of a femtosecond laser to shorter wavelengths. Phase-matched (coherent) addition of the high harmonic x-ray fields emitted by many atoms in the medium is shown. An example electron ionization, propagation, and recollision trajectory is also illustrated.

**Figure 2:** (**A**) Theoretical predictions for scaling of the phase matching cutoffs ($h\nu_{PM} \propto \lambda_L^{1.6-1.7}$) in Ar, Ne and He gases (solid blue, green and red lines respectively) showing that full phase-matching of HHG emission can extend into the multi-keV x-ray region. (**B**) Experimental HHG intensity (linear scale) versus photon energy and pressure at ionization levels close to critical ($\eta_{CR}$), demonstrating that phase matched emission extends to 50 eV in Ar for $\lambda_L$=0.8 μm. (**C** - **E**) Experimental data verifying significant extension of the phase matching cutoffs as $\lambda_L$ increased from 0.8 to 1.3 μm in Ar (**C**), Ne (**D**) and He (**E**). In He, phase matching extends into the water window, using the shortest possible driving laser wavelength. The top panels in **B** - **E** depict the filter transmission curves used to eliminate the fundamental laser beam.

**Figure 3:** (**A**) Experimental macroscopic HHG signal growth versus pressure at photon energies close to the phase matching cutoffs for $\lambda_L$=1.3 μm, demonstrating that higher optimal pressures are needed for longer driving laser wavelengths. In He, the unsaturated signal growth versus pressure for the same propagation lengths as Ne illustrates the decreasing reabsorption in He at high photon energies. (**B**) Theoretical predictions of the required pressure for phase matching for $\lambda_L$=0.8 and 1.3 μm, using a simple semi-analytical model. (**C**) Quantum calculations based on SFA model, also showing that higher pressures are needed to phase match the macroscopic

emission at $\lambda_L$=1.3 μm compared to 0.8 μm in He. (**D**) Evolution of the laser field $E_L(t)$ in He, represented as a temporal overlap of the field at different distances along the waveguide for a pressure of 1300 torr and for a propagation distance of 1 cm. The macroscopic in-phase HHG emission up to the phase matching cutoff is confined within the central cycle of the laser (where there is no phase slip - top inset).

**Figure 4:** (**A**) Experimentally measured flux of phase-matched HHG emission in He, Ne (Ag filter - solid line, Zr filter - dashed line) and Ar, illustrating the extended, sharp cutoffs. (**B**) Raw image of the near-perfect Gaussian soft x-ray beam generated in He, for a broad band of photon energies (200-330 eV), spanning into the water window.

**Figure 5:** (**A**) Critical ionization levels above which full phase matching is not possible, that scale approximately as $\eta_{CR} \propto \lambda_L^{-2}$. (**B**) Predicted laser intensities required to reach the critical ionization level at the peak of the laser pulse on axis (solid line), and when modal averaging is taken into account (dashed line).

**Figure 6:** Scaling of the pressures and lengths required for efficient phase-matched HHG emission at the phase matching cutoffs (dashed red, green, and blue curves for He, Ne and Ar respectively), for laser wavelengths up to $\lambda_L$=6 μm and corresponding phase matching limits spanning to 3 keV. (**A**) Predicted optimal pressure (solid curves) showing quadratic growth with $\lambda_L$ for harmonics generated at ionization levels slightly below critical. (**B**) Corresponding absorption limited medium length (solid curves) for reaching 90% of the asymptotic limit of the macroscopic harmonic signal, determined by reabsorption. Note the substantially longer medium lengths possible in He even at high pressures.

**Figure 7:** Absorption-limited intensity of the phase matched harmonic emission (per unit area and time) at the phase matching cutoffs for $\lambda_L$=0.8-10.0 μm, generated in He, Ne and Ar gases. (**A**) Predicted flux in a bandwidth corresponding to a single harmonic. (**B**) Predicted absorption limited HHG flux in a linewidth of $\lambda/\Delta\lambda$=100 emphasizing the favorable scaling of macroscopic HHG in He in the multi-keV range. The curves in both 3D plots are normalized to the phase-matched HHG emission in each gas at $\lambda_L$=0.8 μm. For high photon energies, group velocity mismatch and magnetic field effects reduce the HHG flux below these predictions (see text for details).

**Figure 8:** (**A**) Ratio, $\xi$, of the amplitude of the quivering electron in the laser field and the speed of light versus $\lambda_L$, showing that relativistic effects do not play significant role under the conditions considered here. (**B**) Schematic of the drift of the electron wavepacket due to the magnetic component of the laser field, along the propagation direction away from the nucleus. (**C**) Squared ratio, $\varepsilon$, of the drift of the wavepacket and its spread versus $\lambda_L$, showing that magnetic field effects will decrease the HHG flux for $\lambda_L$ approaching 6 μm in He.

**TABLE LEGENDS**

**Table 1:** Scaling of the single atom yield with laser wavelength $\lambda_L$ at $h\nu_{PM}$.

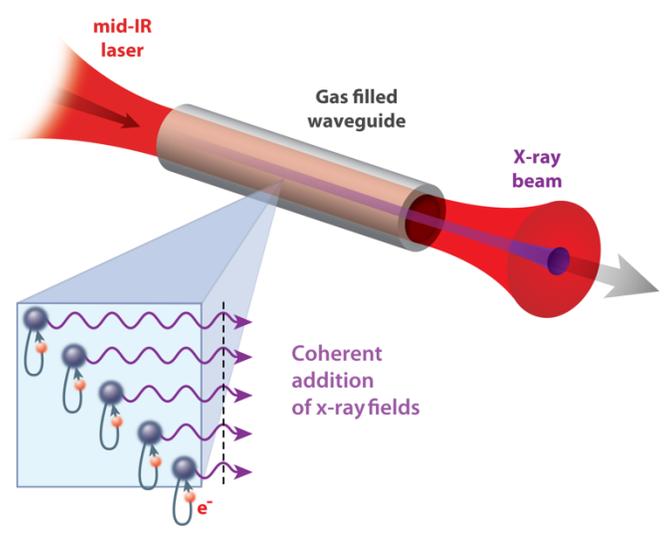

**Figure 1**

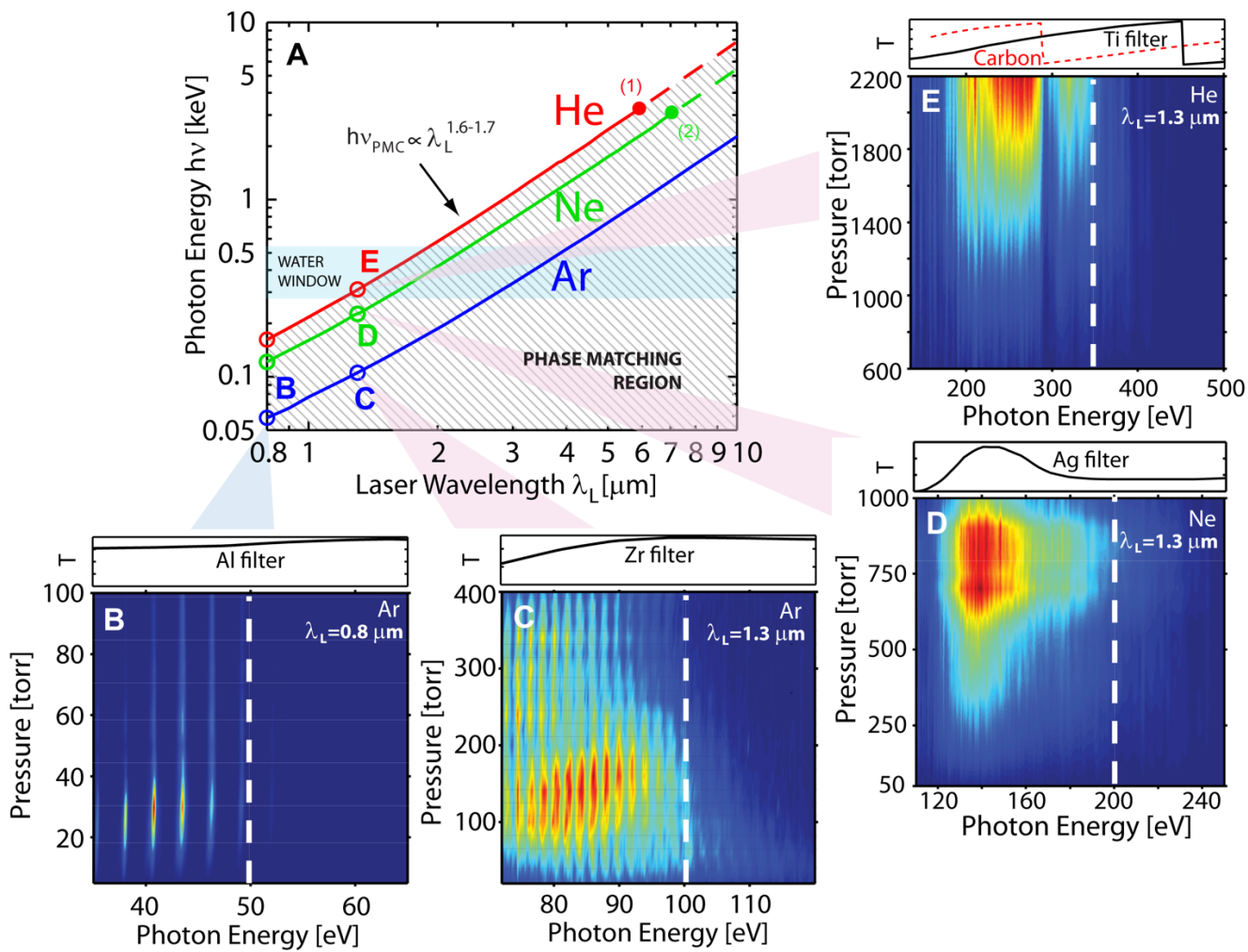

**Figure 2**

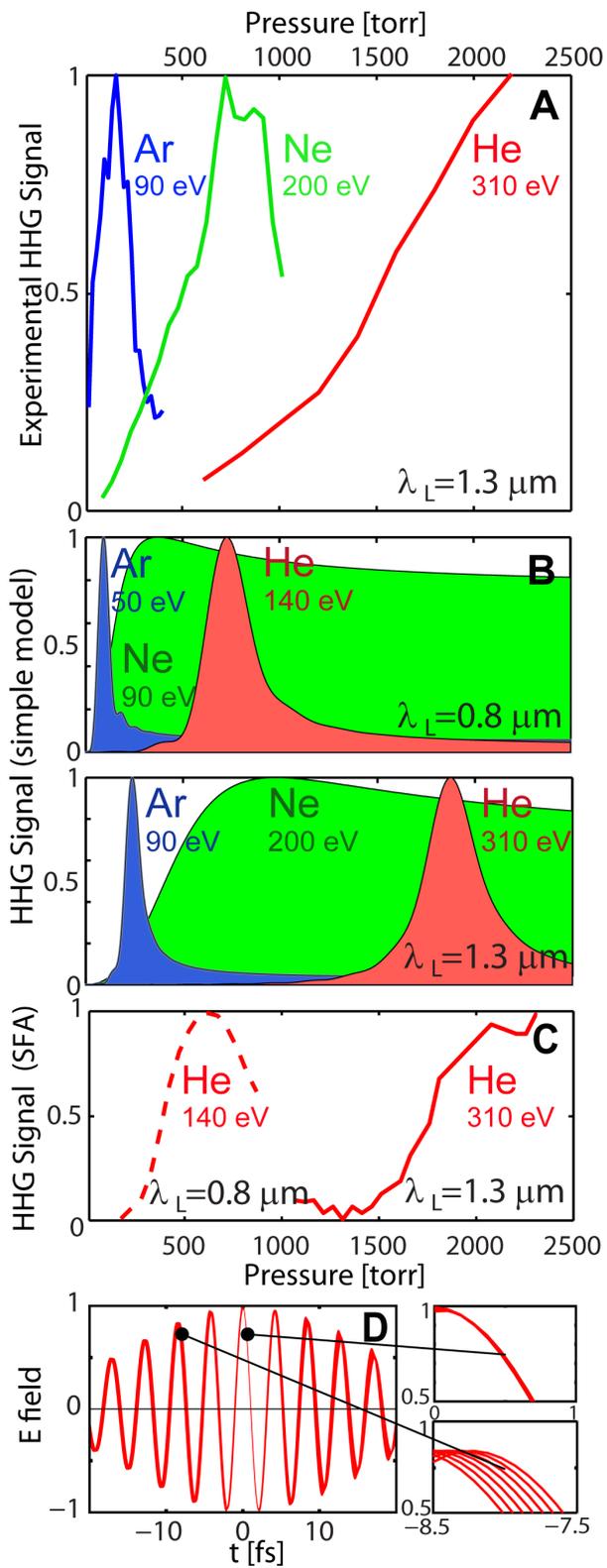

**Figure 3**

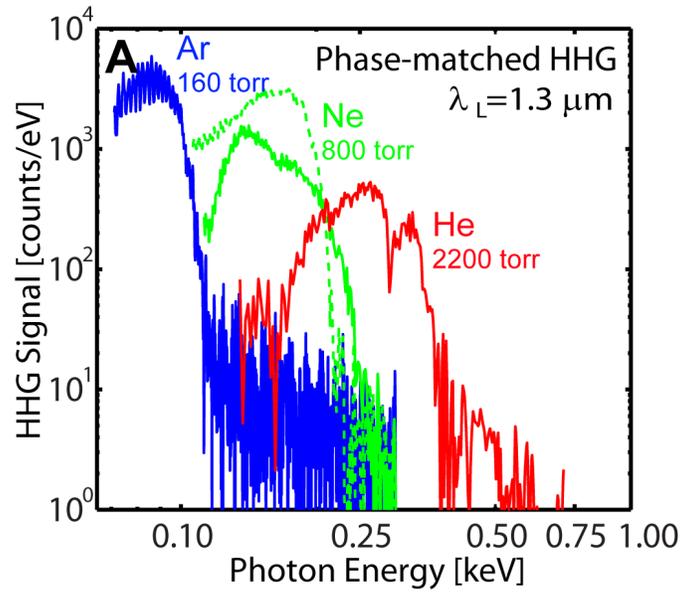

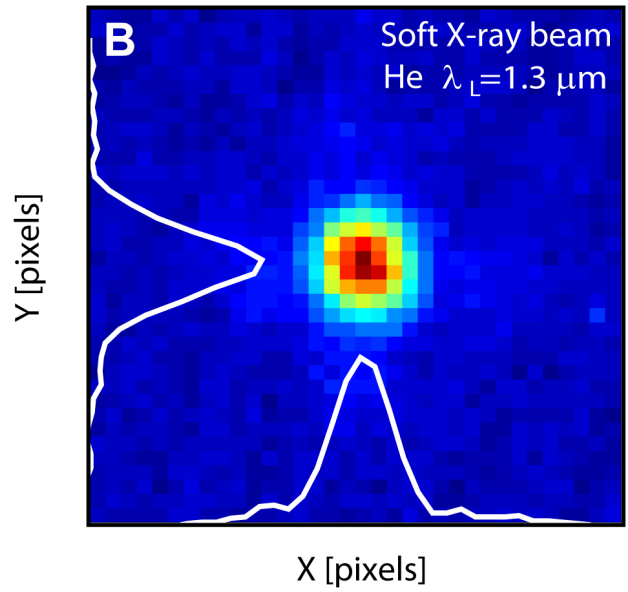

**Figure 4**

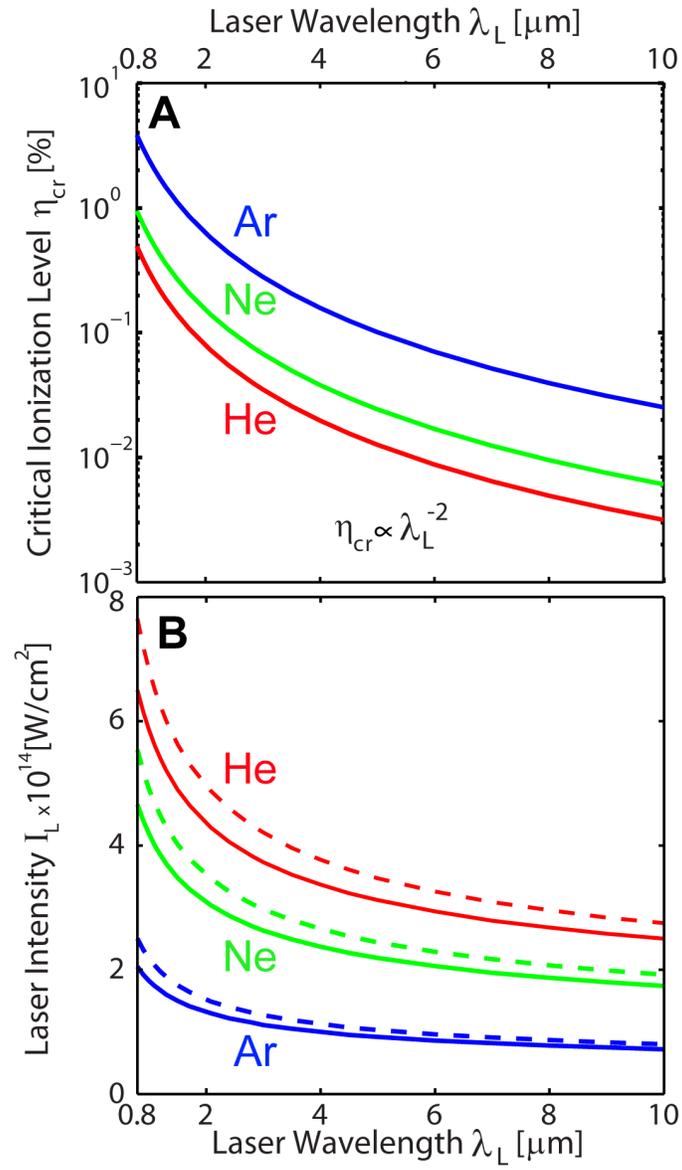

**Figure 5**

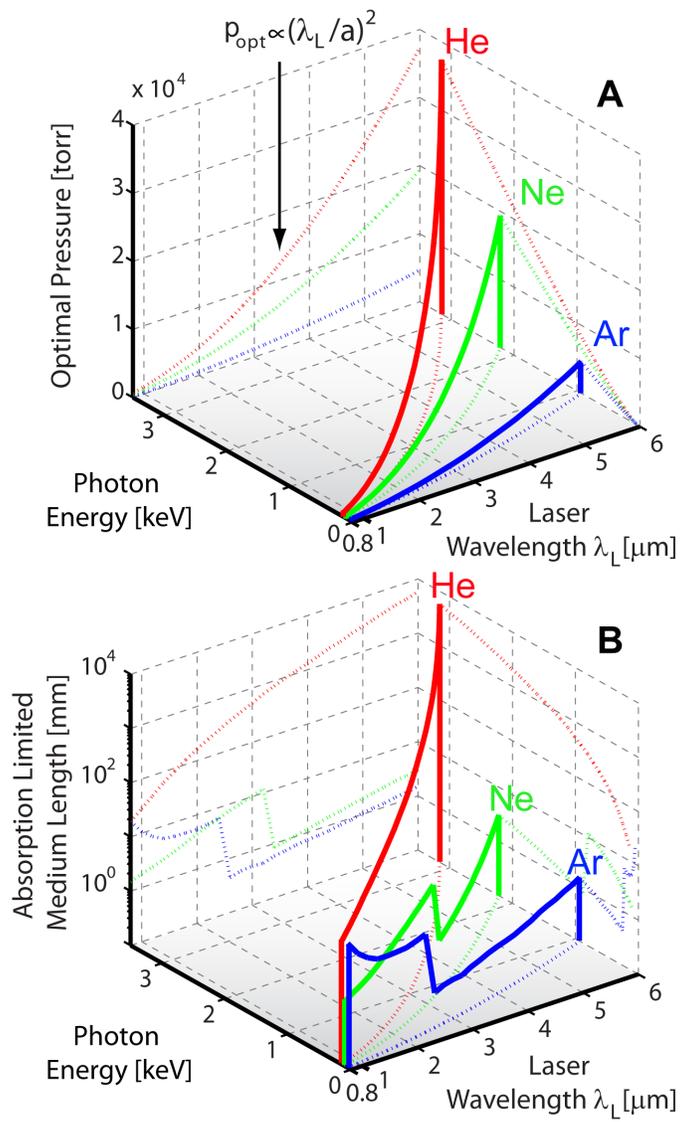

**Figure 6**

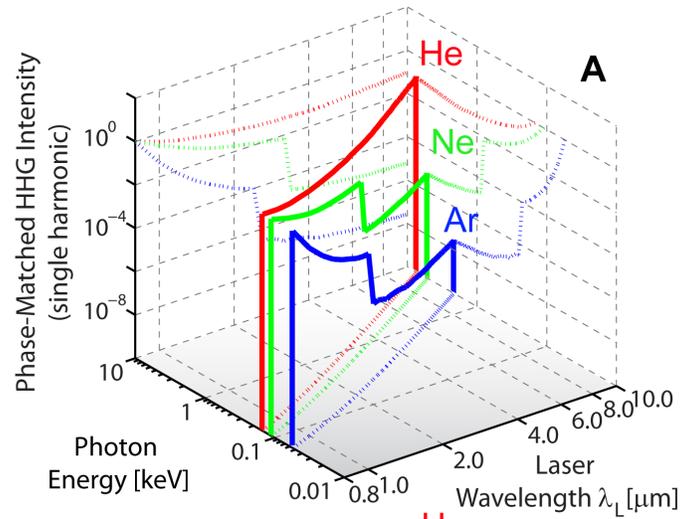

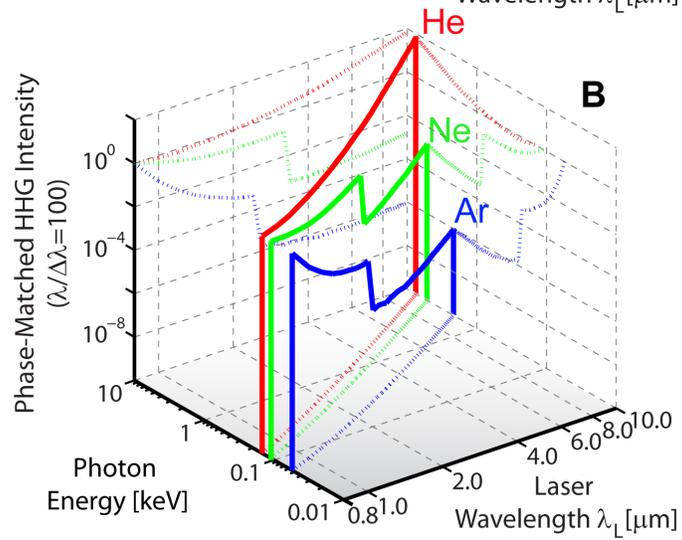

**Figure 7**

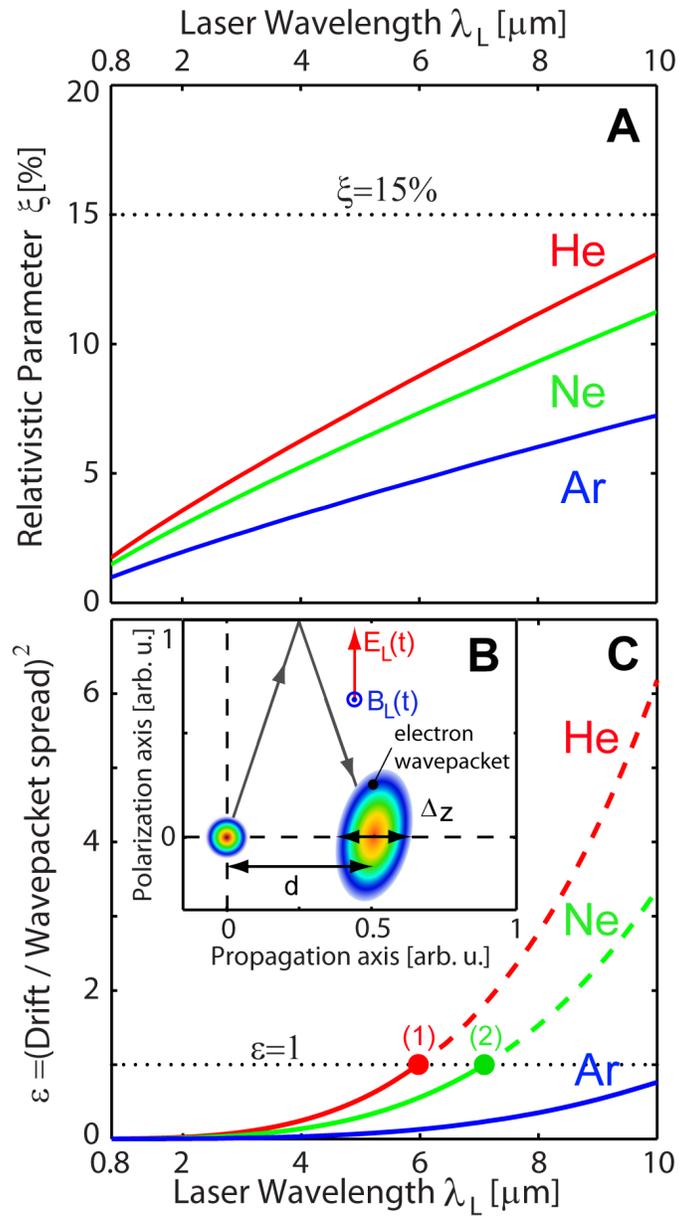

**Figure 8**

| Gas | $x$ | | $y$ | |
| --- | --- | --- | --- | --- |
| | $\Delta E=2E_f$ | $\Delta E=E/100$ | $\Delta E=2E_f$ | $\Delta E=E/100$ |
| He | -9.00 | -6.43 | -14.05 | -11.45 |
| Ne | -8.62 | -6.44 | -13.44 | -11.26 |
| Ar | -8.38 | -6.45 | -12.57 | -11.17 |

**Table 1**